\documentclass[12pt]{article}
\usepackage{geometry}
\usepackage{graphicx,amsfonts,psfrag,fancyhdr,layout,appendix,subfigure}
\usepackage{a4}
\usepackage{graphicx}
\usepackage{epsf}
\usepackage{amsmath}
\usepackage{amssymb}
\usepackage{cite}
\newcommand{\be}{\begin{equation}}
\newcommand{\ee}{\end{equation}}

\newcommand{\Rmnum}[1]{\expandafter\@slowromancap\romannumeral #1@}
\newcommand{\bea}{\begin{eqnarray}}
\newcommand{\eea}{\end{eqnarray}}

\begin{document}
\def\C{{\mathbb{C}}}
\def\R{{\mathbb{R}}}
\def\s{{\mathbb{S}}}
\def\T{{\mathbb{T}}}
\def\Z{{\mathbb{Z}}}
\def\W{{\mathbb{W}}}
\def\Bbb{\mathbb}
\def\BZ{\Bbb Z} \def\BR{\Bbb R}
\def\BW{\Bbb W}
\def\BM{\Bbb M}
\def\BC{\Bbb C} \def\BP{\Bbb P}
\def\CP{\BC\BP}
\begin{titlepage}
\title{Thermodynamic Geometry, Phase Transitions, and the Widom Line}
\author{}
\date{
George Ruppeiner $^a$\footnote{ruppeiner@ncf.edu}, Anurag Sahay $^b$\footnote{anurag@iopb.res.in, Current Address:
Institute of Physics, Bhubaneshwar, India}, Tapobrata Sarkar $^b$\footnote{tapo@iitk.ac.in}, \\ Gautam Sengupta $^b$\footnote{sengupta@iitk.ac.in}
\vskip1.8cm
{\sl $^a$ Division of Natural Sciences,\\
New College of Florida, 5800 Bay Shore Road,\\
Sarasota, Florida 34243-2109
\vskip0.4cm
$^b$ Department of Physics, \\
Indian Institute of Technology,\\
Kanpur 208016, \\
India}}
\maketitle
\abstract{
\noindent
We construct a novel approach, based on thermodynamic geometry, to
characterize first-order phase transitions from a microscopic perspective,
through the scalar curvature in the equilibrium thermodynamic state space.
Our method resolves key theoretical issues in macroscopic thermodynamic
constructs, and furthermore characterizes the Widom line through the maxima
of the correlation length, which is captured by the thermodynamic scalar
curvature. As an illustration of our method, we use it in conjunction with the mean field Van der
Waals equation of state to predict the coexistence curve and the Widom line. 
Where closely applicable, it provides excellent
agreement with experimental data. The universality of our method is indicated by direct calculations from the
NIST database.}
\end{titlepage}
\newpage

Macroscopic properties of matter undergo discontinuous changes along a first-order liquid-gas coexistence curve that culminates in a critical point \cite{callen}, and is extendable
into the supercritical region as the Widom line  \cite{widomline},\cite{stanley},\cite{simeoni} characterized by the locus of points with maximum correlation length $\xi$.
Historically, such phase coexistence curves were modeled by the van der Waals (vdW) equation augmented by the Maxwell ``equal area'' construction.  
This approach lies at the foundation of the modern thermodynamic picture characterizing coexisting phases through equal GibbÕs free energies. 
However, the vdW-Maxwell theory suffers from several unresolved conceptual drawbacks \cite{tisza},\cite{pippard}, and furthermore, an analytic prediction of the Widom line 
from any equation of state is yet unknown.
Here we devise a novel construction to characterize liquid-gas phase transitions based on the continuity of $\xi$ between the phases, with the 
Riemannian geometric thermodynamic scalar curvature $|R| \sim \xi^3$ \cite{ruppeiner}, which also allows, for the first time, a direct computation of the Widom line.
The idea that the correlation lengths of the coexisting phases are equal, and its computational realization through $R$, 
provides a method for predicting the phase coexistence curve when used in conjunction with any equation of state, or experimental data.  
We illustrate this point here with the vdW equation, settling a century old problem 
of thermodynamic computation.  We also determine the location of the Widom line for several fluids both with the vdW equation and with data from the 
NIST database \cite{nist}.
Our results may be used to predict phase behaviour in a wide variety of systems, 
from boiling water to black holes, and promises to have significant impact on diverse areas of physical sciences and engineering.

The key physical idea in our analysis originates from the microscopic picture of first-order 
liquid-gas phase transitions due to Widom \cite{widom}. In this framework, 
spontaneous density fluctuations cause the local 
density $\rho\left({\vec r}\right)$ in a single phase fluid to deviate from the overall mean density $\rho_0$ in some 
complex, time dependent manner. Mathematically, $\rho\left({\vec r}\right) = \rho_0$ corresponds to an intricate contour 
surface that separates two sides with local mean densities $\rho > \rho_0$ and $\rho < \rho_0$. A straight 
line through the fluid intersects this surface at points spaced an average distance $\xi$ apart, 
where $\xi$ is the correlation length characterizing the size of organized structures inside the fluid. 
$\xi$ is generally small in a disorganized system like an ideal gas, but diverges at the critical 
point for real fluids. When a single phase fluid is very near a first-order phase transition, a small amount of a second, minority phase will begin to form. A 
reference point in this single phase fluid typically has local density close to that of either of the two incipient coexisting phases. The typical density difference 
$|\Delta\rho|$ across the contour surface $\rho\left({\vec r}\right)=\rho_0$ thus equals that of the two phases.  Reversing the role of the majority and minority phases
leaves this argument unchanged, with the same $|\Delta\rho|$.  $\xi$ in the single phase plays a similar role in 
anticipating the properties in the two phases since $\xi$ is the thickness of the interface between the two \cite{widom}. This anticipated interface thickness must be the same 
approaching the phase transition with either of the two phases being the majority phase, and the correlation length $\xi$ should thus be the same in the two coexisting 
phases, the condition at the heart of our approach.

\begin{figure}[!t]
\centering
\includegraphics[width=3.5in]{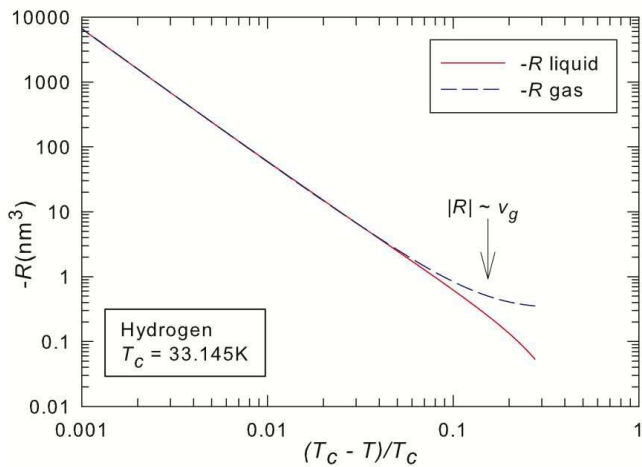}
\caption{-$R$ for the coexisting liquid and gas phases versus $\left(T-T_c\right)/T_c$ for normal Hydrogen calculated with the NIST fluid database, plotted
on the logarithmic scale. At the indicated value, $|R| \sim v_g$, where $v_g$ is the molecular volume in the gas phase. Below this value of
$|R|$, its interpretation as the correlation length loses significance.}
\label{sarkar_fig1}
\end{figure}

For experimental predictability, 
we need an estimate of  $\xi$, allowing a thermodynamic expression for the equality of the 
correlation lengths at the interface. This can be realized using the 
Riemannian geometry of the equilibrium thermodynamic state space of the system through the metric 
$g_{\alpha\beta} = -\frac{1}{k_B}\frac{\partial^2 s} {\partial a^{\alpha}\partial a^{\beta}}$ ($\alpha$, $\beta$ = 1,2), where $k_B$ is Boltzmann's constant, 
and $s$, $a^1$, and $a^2$ denote the entropy, energy, and particle number per unit volume, respectively \cite{ruppeiner}. 
The metric is based on 
Gaussian fluctuation theory whose breakdown takes place when 
the volume of the system is of the order of the Riemann scalar curvature $R$ of the thermodynamic metric. This volume is expected to be the correlation volume $\xi^3$,
leading to the desired connection \cite{ruppeiner}, $|R|\sim \xi^3$. 
Experimental predictions for the coexistence curves of first-order phase transitions can thus be obtained from the equality of $|R|$ 
calculated in the two coexisting phases. We call this the $R$-crossing method. At the critical point, $R$ diverges. In the supercritical region beyond the critical point, 
the locus of the maximum of $|R|$, via $|R|\sim \xi^3$, provides a theoretical prediction of the Widom line.

As a direct test of our proposal, we calculate $R$ for Hydrogen in both phases using the NIST fluid database \cite{nist},\cite{eric} and its program 
REFPROP.  These provide data based on phenomenological equations of state, with fit parameters determined by matching to 
experimental data for fluids.  Results are shown in fig.(1), where agreement between the $R$'s in the two phases is seen to be excellent near the critical point, better 
than $1\%$ in the range $0.96<T/T_c<1$, where $T$ is the temperature and $T_c$ its critical value.  
By contrast, at $T/T_c=0.96$, the molar densities of the coexisting liquid and gas phases differ from each other by a factor of $\sim 3$.

Our $R$-crossing method complements the canonical macroscopic rule for first-order phase transitions, namely
the equality of the molar Gibb's free energies $g$ of the coexisting phases \cite{callen}. 
Applied to the vdW equation however, this macroscopic equality has unresolved conceptual problems. Finding states with equal $g$'s involve contentious issues of
integration along a reversible path through a 
thermodynamically unstable region in the Maxwell construction, or through the critical point in Kahl's approach \cite{kahl}. Such conceptual difficulties, which have
been debated for over a century, are entirely bypassed in our construction. Further, our method naturally contains a measure of its 
limit  of applicability, since for the construction to be effective, $\xi^3$ should be large enough to encompass a number of atoms adequate for a thermodynamic approach to be reasonable. 
This limits us to a regime not too far from the critical point. 
We find that the $R$-crossing method retains its viability down to volume regimes containing but about a single molecule. 

\begin{figure}[!t]
\centering
\includegraphics[width=3.0in,height=2.5in]{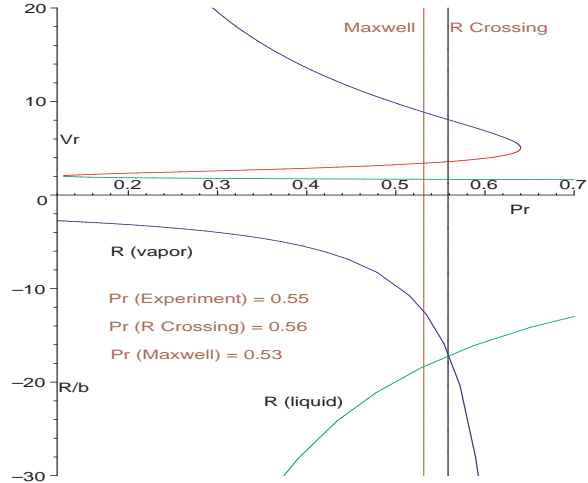}
\caption{$R$ vs $p_r$ along an isotherm of Helium with $t_r=0.86$ in the lower half and $v_r$ vs $p_r$ along the same $t_r$ in the upper half 
(where the $v_r$ values have been multiplied by a factor of 3).
The blue and green curves represent the stable branches, and the red curve is the unstable branch. We mark by ``$R$-crossing'' 
the $p_r$ where the $R$'s of the liquid and gas phases become
equal (with $c_v=1.5$ and $1.2$ on the gas and liquid sides respectively). The line labeled 
``Maxwell'' represents the corresponding $p_r$ from Maxwell's construction.}
\label{sarkar_fig2}
\end{figure}

As a simple theoretical example, we apply the $R$-crossing method to the 
universal vdW equation in its reduced form,
\begin{equation}
p_r = \frac{8t_r}{3v_r - 1} - \frac{3}{v_r^2},
\label{vdw}
\end{equation}
where $p_r = P/P_c$, $t_r = T/T_c$, $v_r = v/v_c$ and $P$ and $v$ are the pressure and molar volume, with 
the subscript $c$ denoting their critical values. The critical quantities are known to be related to the vdW constants $a$ and $b$ by 
$P_c = \frac{a}{27b^2}$, $T_c = \frac{8a}{27k_BT}$ and $v_c = 3b$. The Maxwell equal-area construction
yields the limiting slope of the coexistence curve $dp_r/dt_r = 4$, independent of the fluid and its heat capacity. 
Our $R$-crossing method inherits the same limiting slope here. This number is closely followed only
by Helium and Hydrogen, for which this example is expected to be maximally effective.  
$R$ can be calculated here
via the thermodynamic metric using standard formulae \cite{ruppeiner} and gives $R=A \cdot B$, where
\begin{equation}
A = -\frac{b}{3}\frac{3v_r - 1}{c_v\left(p_rv_r^3 - 3v_r + 2\right)^2},
\label{r1}
\end{equation}
and 
\begin{eqnarray}
B &=& c_v\left(p_r^2v_r^5 - 9p_rv_r^4 + 12p_rv_r^3 - 27v_r^2 - p_rv_r^2  + 27v_r - 3\right)\nonumber\\
&~& +18v_r\left(p_rv_r^3 + 1\right)
\label{r2}
\end{eqnarray}

\noindent
where $c_v$ is the dimensionless molecular specific heat at constant volume (assumed constant, though 
possibly different in the liquid and gas phases) and $b$ plays no role in our subsequent analysis. 

For vdW isotherms with given $t_r<1$, substituting $p_r$ from eq.(\ref{vdw}) into eqs.(\ref{r1}) and (\ref{r2}) results in
two physical branches for $R$, corresponding to the liquid and gas phases (see color coded fig.(2)), with $R$ diverging at the end points. 
The value of $p_r$ where the $R$ values are equal (i.e they cross) is interpreted as the reduced saturation pressure corresponding to $t_r$.
For the cases we consider here, $c_v$ on the gas side is taken as 1.5, the ideal gas value. 
On the liquid side, we have chosen average values determined from NIST data \cite{nist}, over the range of  temperatures 
that we are interested in. Equivalently, for vdW isobars, the $R$-crossing method can
be used to predict the saturation temperature.
\begin{figure}[t!]
\begin{minipage}[b]{0.5\linewidth}
\includegraphics[width=2.8in]{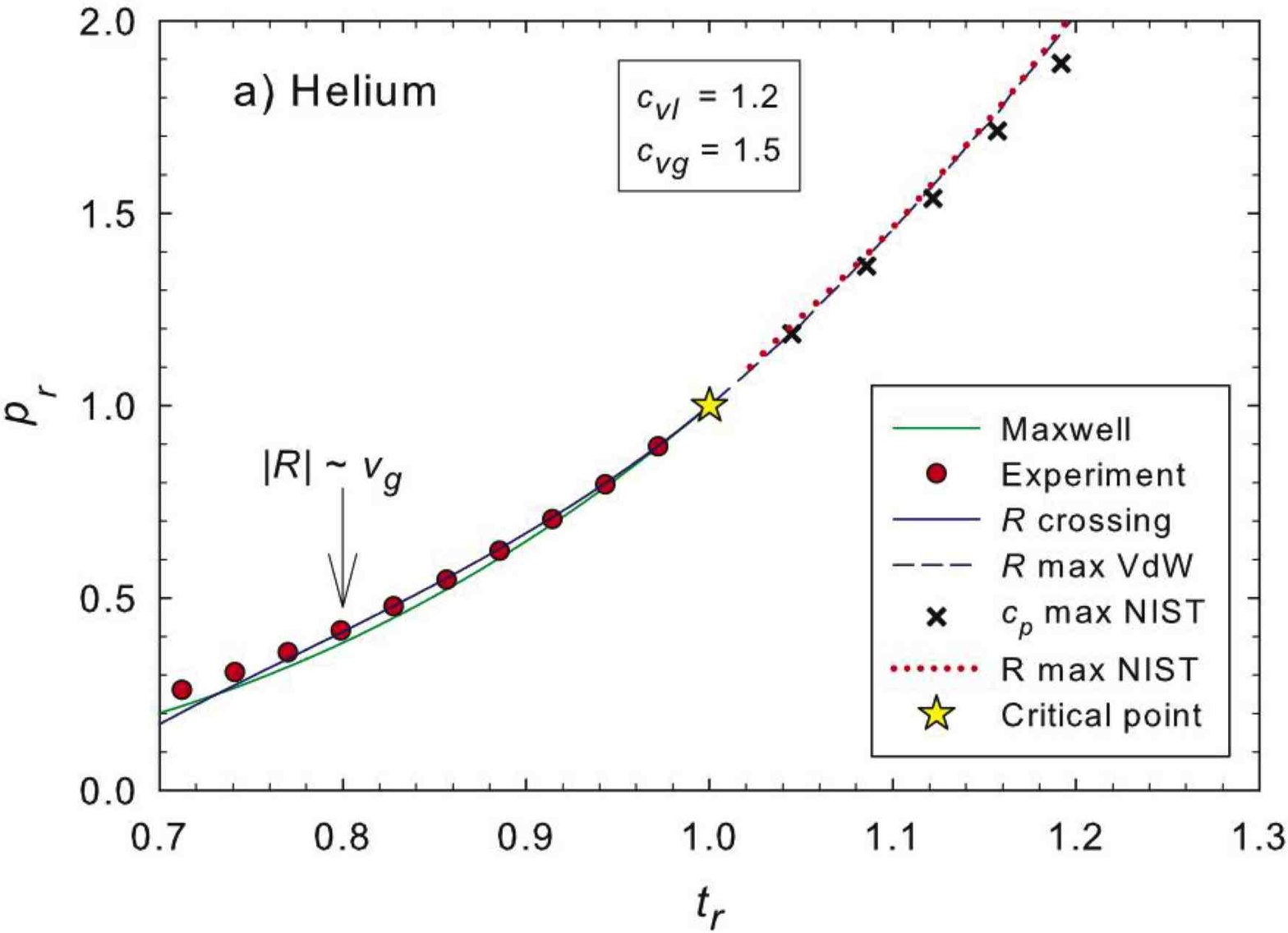}
\end{minipage}
\hspace{0.2cm}
\begin{minipage}[b]{0.5\linewidth}
\includegraphics[width=2.8in]{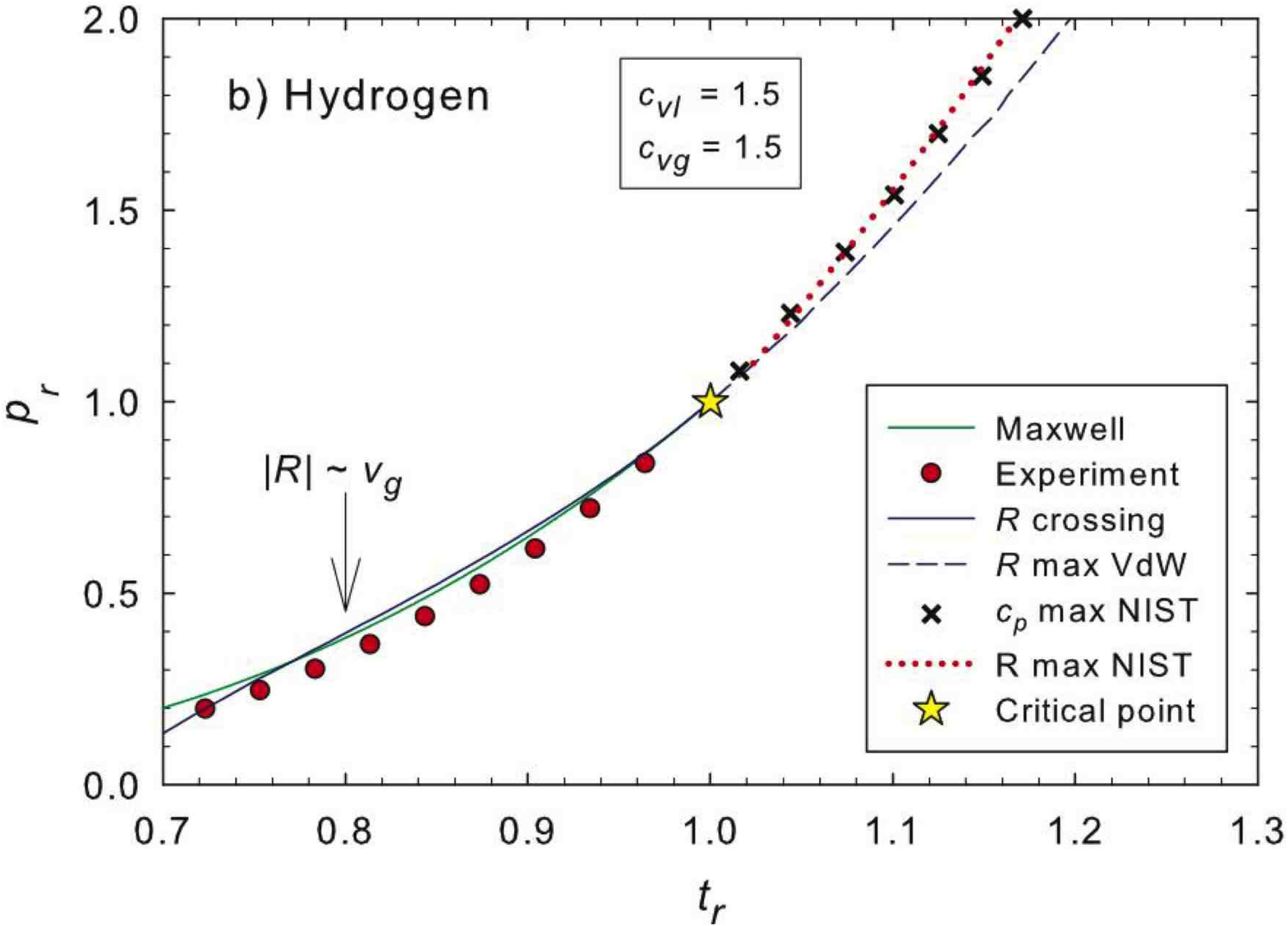}
\end{minipage}
\caption{ Phase coexistence and the Widom line for Helium ($T_c = 5.19$K, $P_c = 2.26$ bars) on the left
and Hydrogen ($T_c = 33.19$K, $P_c=13.30$bars) on the right.  The coexistence curve is calculated from vdW with the 
Maxwell equal-area construction and with $R$-crossing, and compared with experimental data from NIST \cite{nist}.  The Widom line is 
calculated by finding the locus of maximum values of $|R|$ both with vdW and from NIST data.
 We compare with
the maximum values of $c_p$ from experimental data (NIST). The liquid and gas heat capacities $c_{vl}$ and 
$c_{vg}$ are indicated for vdW. In the supercritical region, we use $c_{vg}$}
\label{sarkar_fig3}
\end{figure}

\begin{table}[ht]
\begin{minipage}[b]{0.4\linewidth}\centering
\begin{tabular}{|c| c| c| c|}
\hline
$P/P_c$ & $T_R^{sat}$(vdW) & $T_{ex}^{sat}$ & $|R|/v_g$\\ \hline \hline 
0.4 &  36.04 & 37.97  & 0.60 \\
0.5 & 37.68 & 39.41  & 1.16 \\
0.6 & 39.24 & 40.66 & 2.36 \\
0.8 & 42.09 &  42.76 & 15.35\\
0.9 & 43.33 & 43.66 &  76.48\\
\hline\hline
0.4 &  122.89 & 129.16  & 0.57 \\
0.5 & 128.19 & 133.93  & 1.11 \\
0.6 & 133.34 & 138.07 & 2.23 \\
0.8 & 142.70 &  145.00 & 14.29\\
0.9 & 146.85 & 147.98 &  72.49\\
\hline
\end{tabular}
\end{minipage}
\begin{minipage}[b]{0.7\linewidth}\centering
\begin{tabular}{|c|c|c|c|}
\hline
$P/P_c$ & $T_R^{W}$(vdW) & $T_R^W$(NIST) & $T^W_{ex}$ \\ \hline \hline 
1.1 & 45.56 & 45.25 & 45.26 \\
1.2 & 46.57 & 45.95 & 46.01 \\
1.4 & 48.43 & 47.26 & 47.39 \\
1.6 & 50.15 & 48.50 & 48.64 \\
2.0 & 53.26 & 50.83 & 50.79 \\
\hline\hline
1.1 & 154.32 & 153.15 & 153.21 \\
1.2 & 157.74 & 155.47 & 155.60 \\
1.4 & 164.04 & 159.72 & 160.00 \\
1.6 & 169.84 & 163.69 & 163.89 \\
2.0 & 180.41 & 170.96 & 170.49 \\
\hline
\end{tabular}
\end{minipage}
\label{tableone}
\caption{Saturation temperatures on the left and Widom line temperatures on the right (in Kelvins) for Neon ($T_c=44.49$ K, $P_c=26.79$ bars) on the upper part
and Argon ($T_c=150.69$ K, $P_c = 48.63$ bars) on the lower part. $T_R^{sat}$(vdW) is the prediction of the saturation temperature from the $R$-crossing method, using
the vdW equation, and is compared with experimental values from NIST. Corresponding values of $|R|/v_g$ are also shown to indicate the validity of our method. 
Widom line predictions from the $R$-maximization method are obtained both from vdW with $c_v = 1.5$ ($T_R^W$(vdW)),
and from NIST ($T_R^W$(NIST)). We have also shown the prediction of the Widom line obtained from maximising $c_p$ as $T^W_{ex}$.}
\end{table}

In the supercritical region, isobaric $|R|$ exhibits a local maximum with respect to $t_r$, whose locus is naturally interpreted as 
the Widom line, signifying a crossover for certain dynamical fluid properties from gas like on the low pressure side to liquid like 
on the high pressure side \cite{widomline},\cite{stanley},\cite{simeoni}. 
We can calculate the Widom line as per its definition through $|R| \sim \xi^3$, free from
the theoretical difficulty of characterizing it via the maximum of the specific heat $c_p$ as is conventional in the literature \cite{stanley}.
To find maxima for $|R|$ and $c_p$, we search along isobars. 

A natural estimate for the validity of our analysis for vdW is offered by the dimensionless quantity $|R|/v_g$, where $v_g$ is the coexistence molecular 
volume in the gas phase.  $|R| \geq v_g$ implies that $\xi^3$ is greater
than a molecular volume, and we are in a regime where our analysis is strictly valid. We find that for vdW,
this restricts us to $t_r \gtrsim 0.8$ along the coexistence curve (a value indicated in fig.(3) and table (1)), and to 
$p_r \lesssim 10$, in the supercritical region.

Figure (3) summarizes our results for Helium and Hydrogen. Table (1) supplements these for Neon and Argon. 
From fig.(3), it can be seen that the $R$-crossing method, in conjunction with vdW, predicts excellent results within its range of applicability. 
Away from criticality, deviation from data is also due to the mean field nature of the vdW equation of state. Direct application of the $R$-maximization
method using NIST data in the supercritical regime shows striking agreement with experimental $c_p$ maximum values even far from the scaling region. 

In conclusion, we have constructed a novel geometrical technique to characterize liquid-gas phase transitions from a microscopic perspective, through
the thermodynamic scalar curvature $R$.  When applied in conjunction with the vdW equation, this bypasses theoretical issues a century old with the Maxwell 
equal area construction and its variants.  Our technique generalizes to any phenomenological equation of state, including those obtained as multi parameter fits to 
experimental data.  This analysis further provides the first direct theoretical construction of the Widom line, without using any ad hoc thermodynamic response function.

 Our method unifies concepts in Riemannian geometry, thermodynamics, phase transitions, critical and supercritical phenomena.  
 Although we have primarily applied our technique to liquid-gas phase transitions, the method should be universally applicable to any first-order phase transition. 
 This makes it of crucial significance to a diverse range of disciplines in physical, chemical and biological sciences, and engineering. It further generalises even to 
 gravitational systems like anti de-Sitter black holes which also appear to exhibit liquid-gas like first-order phase transitions \cite{sss}. \\

\vspace{0.2in} 
\begin{center}
{\bf Acknowledgements}
 \end{center}
We thank Steven Shipman and Helge May for valuable input, and Eric Lemmon for programming $R$ into REFPROP 9.01, 
allowing us to verify the numbers in Figure 1 and readily compute Widom lines with $R$. TS thanks the Saha Institute of Nuclear Physics, Kolkata, India for 
its hospitality where a part of this work was completed.\\

\end{document}